\documentclass[aps,pre,twocolumn,superscriptaddress,showpacs,showkeys ]{revtex4-2}

\usepackage{graphicx}                   
\usepackage{hyperref}                   
\usepackage{amsmath,amssymb}            
\usepackage{epstopdf}
\usepackage{subcaption}					
\usepackage{color}
\definecolor{orange}{rgb}{1,0.5,0}
\definecolor{brown}{rgb}{0.65, 0.16, 0.16}
\definecolor{phlox}{rgb}{0.87, 0.0, 1.0}

\begin{document}

\title{Self-similar but not conformally invariant traces obtained by modified Loewner forces}

\author{S. Tizdast}
\affiliation{Department of Physics, University of Mohaghegh Ardabili, P.O. Box 179, Ardabil, Iran}
\email{morteza.nattagh@gmail.com}

\author{Z. Ebadi}
\affiliation{Department of Physics, University of Mohaghegh Ardabili, P.O. Box 179, Ardabil, Iran}

\author{J. Cheraghalizadeh}
\affiliation{Department of Physics, University of Mohaghegh Ardabili, P.O. Box 179, Ardabil, Iran}

\author{M. N. Najafi}
\affiliation{Department of Physics, University of Mohaghegh Ardabili, P.O. Box 179, Ardabil, Iran}

\author{Jos\'e S. Andrade}
\affiliation{Departamento de F\'{i}sica, Universidade Federal do Cear\'{a}, 60451-970 Fortaleza, Brazil}

\author{Hans J Herrmann}
\affiliation{Departamento de F\'{i}sica, Universidade Federal do Cear\'{a}, 60451-970 Fortaleza, Brazil}
\affiliation{PMMH, ESPCI, 17 quai St. Bernard, 75005 Paris, France}

\begin{abstract}
The two-dimensional Loewner exploration process is generalized to the case where the random force is self-similar with positively correlated increments. We model this random force by a fractional Brownian motion with Hurst exponent $H\geq \frac{1}{2}\equiv H_{\text{BM}}$, where $H_{\text{BM}}$ stands for the one-dimensional Brownian motion. By manipulating the deterministic force, we design a scale-invariant equation describing self-similar traces which lack conformal invariance. The model is investigated in terms of the ``input diffusivity parameter'' $\kappa$, which coincides with the one of the ordinary Schramm-Loewner evolution (SLE) at $H=H_{\text{BM}}$. In our numerical investigation, we focus on the scaling properties of the traces generated for $\kappa=2,3$, $\kappa=4$ and $\kappa=6,8$ as the representatives, respectively, of the dilute phase, the transition point and the dense phase of the ordinary SLE. The resulting traces are shown to be scale-invariant. Using two equivalent schemes, we extract the fractal dimension, $D_f(H)$, of the traces which decrease monotonically with increasing $H$, reaching $D_f=1$ at $H=1$ for all $\kappa$ values. The left passage probability (LPP) test demonstrates that, for $H$ values not far from the uncorrelated case (small $\epsilon_H\equiv \frac{H-H_{\text{BM}}}{H_{\text{BM}}}$) the prediction of the ordinary SLE is applicable with an effective diffusivity parameter $\kappa_{\text{eff}}$. Not surprisingly, the $\kappa_{\text{eff}}$'s do not fulfill the prediction of SLE for the relation between $D_f(H)$ and the diffusivity parameter.
\end{abstract}


\maketitle
\section{Introduction}
The Loewner exploration process for generating conformal invariant (CI) random traces helps to uncover statistical properties of loopless paths of two-dimensional critical models with conformal invariance, and therefore makes a bridge between these models and self-similar stochastic processes. In an equivalent backward process, the Schramm-Loewner evolution (SLE) theory maps the 2D stochastic traces to a one-dimensional Brownian motion (1DBM), organizing the traces into universality classes distinguished by one single parameter~\cite{cardy2005sle,schramm2000scaling}. The representative of each class is the proportionality coefficient ($\sqrt{\kappa}$) of the relation between the driving function of the SLE map and the 1DBM, {\it i.e.}, the variance of the driving function is proportional to time (as the parametrization of the stochastic graphs), with the proportionality coefficient $\kappa$. The importance of this theory is due to the fact that many properties of its traces are known. Examples of the predictions of SLE are the left passage probability (LPP)~\cite{najafi2013left,cheraghalizadeh2019correlation}, the Fokker-Planck equation~\cite{najafi2015fokker}, the winding angle (WA) statistics~\cite{boffetta2008winding,najafi2018schramm}, the crossing probability~\cite{cardy2005sle}, the relation between the fractal dimension and $\kappa$~\cite{cardy2005sle,Najafi2012Observation}, locality for $\kappa=6$ and restriction for $\kappa=\frac{8}{3}$~\cite{cardy2005sle}.\\
Despite of its power in characterizing models, the SLE theory is applicable only to systems fulfilling conformal invariance. There are, however, some circumstances under which one is interested in traces which are not conformally invariant. The SLE driven by L\'evy flights is an example, where the driving function is an addition of a 1DBM and a L\'evy process~\cite{oikonomou2008global}. Using the techniques described in Ref.~\cite{oikonomou2008global}, the "suboriented" SLE was developed~\cite{nezhadhaghighi2011first}, by considering time as a non-decreasing stochastic parameter ({\it i.e.},  subordinating the process by the inverse time). Another example concerns anisotropic systems for which application of the SLE theory on anisotropic traces gives a power-law driving function~\cite{credidio2016stochastic}, the exponent of which depends on the strength of anisotropy. This motivates us to analyze a more general Loewner exploration process, where the stochastic ``\textit{force}'' follows a power-law with time. At the same time, one should also generalize the ``deterministic force'' in order to recover the scale invariance of the governing equations, giving rise to self-similar traces. This is the aim of our paper, in which we use fractional Brownian motion (FBM) with Hurst exponent $H$ as the time series for the stochastic force (the driving function).\\

The paper is organized as follows: In Section~\ref{SEC:the loewner} we introduce the Loewner process as backward SLE equations (known also as the Langevin equations for SLE) and generalize it in a systematic way. After introducing the fractional Brownian motion (FBM), we also describe our statistical tools for characterizing the random traces, {\it i.e.}, LPP. We numerically simulate the traces in Section~\ref{SEC:results} and analyze the results. Conclusions are presented in Section~\ref{SEC:conclusion}.

\section{The Loewner exploration process}\label{SEC:the loewner}
2D extended traces, such as the domain-walls of 2D random systems or the trace of random walkers, can often be viewed as an \textit{exploration processes}~\cite{cardy2005sle}. For conformally invariant models, the global properties of self-avoiding domain-walls are well understood thanks to the SLE theory, provided that they have domain Markov property~\cite{schramm2005basic,najafi2013left}. The corresponding Loewner exploration process is defined as the exploration process governed by the backward SLE equation~\cite{najafi2015fokker}, to be described in the next section.
\subsection{SLE and Loewner exploration process}
Let us give a brief introduction to (chordal) SLE, which aims to describe 2D random traces ({\it e.g.}, the domain-walls in 2D conformally invariant statistical models) by parametrizing them with \textit{time} $t$, mapping them to a dynamical process. Let us denote the trace by $\gamma_t:t\rightarrow z_t=x_t+iy_t$, which is a non-intersecting path in the upper half plane $ S = \left\{ {z \in C,{\mathop{\rm Im}\nolimits} z > 0} \right\}$. The hull $K_t$ is defined as the set of points which are located exactly on $\gamma_t$, or are disconnected from infinity by the trace. Then SLE is defined by unique conformal maps (parametrized by $t$) $g_t(z):S_t\rightarrow S$, where $S_t:=S\backslash K_t$ is the complement of $K_t$, {\it i.e.}, $g_t(z)$ \textit{uniformizes} the trace by sending it to the real axis. $g_t(z)$ is the solution of the stochastic Loewner equation~\cite{cardy2005sle,bernard2007inverse,smirnov2001critical}
\begin{equation}
\partial_{t}g_{t}(z)=\frac{2}{g_{t}(z)-\xi(t)},
\label{Eq:mainSLE}
\end{equation}
with the initial condition $ {g_{t = 0}}(z) = z $, where $\xi (t)$ is a real valued function called the driving function. In the hydrodynamical normalization $ {g_t}(z) = z + \frac{{2t}}{z} + o(\frac{1}{{{z^2}}}) $  as $ z \to \infty  $, which fixes $g_t(z)$ according to the Riemann mapping theorem~\cite{bell2015cauchy}. For fixed $z$, $g_t(z)$ is well-defined up to time $\tau_z$ for which $g_t(z)=\xi_t$, and the tip of $\gamma_t$ is mapped to $\xi_t$ on the real axis. Then the hull (up to time $t$) is formally defined as $ K_t \equiv \overline {\{ z \in S:{\tau_z}\le t\}}$. Schramm~\cite{cardy2005sle,schramm2000scaling} proved that for conformally invariant systems $\xi_t$ is a real valued function proportional to a one dimensional (1D) Brownian motion (BM, denoted by $B_t$), {\it i.e.}, $\xi_t=\sqrt{\kappa}B_t$ where $ \kappa $ is a proportionality coefficient known as the diffusivity parameter. \\

Many two dimensional critical statistical models are classified using this theory. For example loop-erased random walks and Abelian sandpiles correspond to $\kappa=2$, the critical Ising model corresponds to $\kappa=3$, the harmonic explorer and the Gaussian free field correspond to $\kappa=4$, critical percolation corresponds to $\kappa=6$, and the uniform spanning trees correspond to $\kappa=8$~\cite{cardy2005sle}.  $\kappa<4$ ($\kappa>4$) belong to the dilute (dense) phase, while $\kappa=4$ is the transition point, and $\kappa=8$ is a space filling trace.\\

The Markov property, stationarity and continuity of the SLE process follow from the properties of the 1DBM: ({\it i}) the domain Markov property is due to the fact that for $s>t>t'$, $\xi_s-\xi_t$ is independent of $\xi_{t'}$; ({\it ii}) the stationarity has its origin in the fact that $\xi_s-\xi_t$ only depends on $s-t$, and ({\it iii}) the continuity is expected since $\lim_{s\rightarrow t}(\xi_s-\xi_t)=0$ for all values of $t$~\cite{schramm2000scaling,cardy2005sle,Najafi2012Observation,Bauer2003Conformal,najafi2015fokker,najafi2013left}. Notice that $ \xi_{t} \stackrel{d}{=} \xi_{-t} \stackrel{d}{=} -\xi_{t} $ is an important symmetry relation for uncovering the properties of the driving function~\cite{Bauer2003Conformal}, where $\stackrel{d}{=}$ means the equality of the distributions of stochastic processes. A deep connection between SLE and conformal field theory (CFT)~\cite{credidio2016stochastic,bauer2002slekappa,bauer2003sle,cardy2005sle,najafi2015observation,najafi2015fokker,najafi2013left} stimulated many analytical and numerical studies for obtaining properties of 2D conformally invariant models~\cite{bernard2006conformal,Najafi_2018,bernard2007inverse,saberi2008conformal}. This correspondence is made via the relation between the central charge $c$ in CFT and the diffusivity parameter $\kappa $ in SLE, {\it i.e.}, $c=(6-\kappa)(3\kappa-8)/(2\kappa)$. \\

We sometimes need to make stochastic curves $(x_t,y_t)$ from an already known BM time series, {\it i.e.}, the inverse of the above process. This is possible defining the backward SLE equation. Let $I_t$ be a conformal homeomorphism from $ S \to {S_t} $ satisfying the equation
\begin{equation}
\partial_{t}I_{t}(w)=\frac{-2}{I_{t}(w)-\xi(t)},
\label{Eq:BmainSLE}
\end{equation}
normalized in such a way as to satisfy $ {I_t}(w) = w - \frac{{2t}}{w} + o(\frac{1}{{{w^2}}})$ as $ w \to \infty$. It was shown that the probability distribution of $ I_{t} $ is the same as that of $ g_t^{ - 1} $~\cite{oikonomou2008global,najafi2015fokker}, {\it i.e.}, $ {I_t}(\omega ) - {\xi _t} \stackrel{d}{=} g_t^{ - 1}(\omega  + {\xi _t}) \stackrel{d}{=} h_t^{ - 1}(\omega ) $ where $ {h_t}(z) $ is the shifted conformal map, {\it i.e.}, $ {h_t}(z) = {g_t}(z) - {\xi _t} $. In particular, the tip of the SLE trace can be obtained by $ {\Gamma _T} = {\lim _{\omega  \to 0}}g_T^{ - 1}(\omega  + {\xi _T}) = {\lim _{\omega  \to 0}}h_T^{ - 1}(\omega ) $. Therefore, using the backward equation of the SLE we find the trajectory of the tip of the trace $ {z_t} = {x_t} + i{y_t} $ (notice that Re($ {\Gamma _t} $) and Im($ {\Gamma _t} $) have the same joint distribution as $ {x_t} $ and $ {y_t} $, respectively),
\begin{equation}
\begin{split}
& \frac{\text{d}}{\text{d}t}x_t=-\frac{2x_t}{x_t^2+y_t^2}-\frac{\text{d}}{\text{d}t}\xi_t\\
& \frac{\text{d}}{\text{d}t}y_t=\frac{2y_t}{x_t^2+y_t^2},
\end{split}
\label{Eq:SLElangavin}
\end{equation}
with the initial values $x_0=u$ and $y_0=v$ in which $w=u+iv$ is the initial point.\\

We note that $\xi(\lambda^2t)\stackrel{d}{=}\lambda\xi(t)$ ($\lambda$ being a non-zero positive scaling parameter), which implies special scaling properties of SLE, and helps understanding the space-time structure of the exploration processes. The scaling relation in Eq.~(\ref{Eq:mainSLE}) and consequently Eq.~(\ref{Eq:SLElangavin}) implies that $g_{\lambda^2t}(\lambda z)\stackrel{d}{=} g_t(z)$, and $x_{\lambda^2t}\stackrel{d}{=} \lambda x_t$ and $y_{\lambda^2t}\stackrel{d}{=} \lambda y_t$. One can directly examine this by applying $t\rightarrow \lambda^2t$ and $z\rightarrow\lambda^{\alpha} z$ (in which $\alpha$ is an exponent to be determined). Then the SLE equation a
\begin{equation}
\frac{1}{\lambda^2}\partial_{t}g_{\lambda^2t}(\lambda^{\alpha} z)=\frac{2}{g_{\lambda^2t}(\lambda^{\alpha} z)-\xi(\lambda^2t)},
\end{equation}
which can only be satisfied if $\alpha=1$ and $g_{\lambda^2t}(\lambda z)=\lambda g_t(z)$, as mentioned above, which is also expected from the very beginning considering the initial condition $g_0(z)=z$. This exploration process can be re-phrased in terms of a self-avoiding walker subjected to random forces, which provides another perspective on the Loewner evolution. Considering $x_t$ and $y_t$ as the components of this random exploration process, the right hand side of Eq.~(\ref{Eq:SLElangavin}) plays the role of a stochastic \textit{force}, called the Loewner force, and is represented by $\vec{f}_L$. This force, that determines the global properties of the model at hand, has two components, $\vec{f}_L\equiv \vec{f}_d+\vec{f}_r$, which are a deterministic force $\vec{f}_d\equiv \frac{2}{r^2}(-x,y)$ , where $r\equiv (x^2+y^2)^{1/2}$ is the distance from the origin, and a random force $\vec{f}_r\equiv -\left( \frac{\text{d}}{\text{d}t}\zeta_t\right) \hat{x}$, that brings stochasticity into the problem. Then, for an exploration process driven by the Loewner force, the trace becomes (isotropic) self-similar with an already known fractal dimension $D_f$, provided that $\zeta_t$ is proportional to a 1DBM, {\it i.e.}, $\zeta_t=\sqrt{\kappa}B_t$, ($D_f=1+\frac{\kappa}{8}$). A very important property of the exploration process driven by the Loewner force is conformal invariance, meaning that the probability measure of the traces is invariant under conformal transformations.
 
\subsection{Exploration process produced by modified Loewner forces}\label{SEC:modifed}
The natural question arising here concerns the possible modifications of the stochastic force $\vec{f}$ on the right hand-side of Eq.~(\ref{Eq:SLElangavin}) in such a way that the trajectory of the random traces, previously introduced, remains self-similar. Changing both the deterministic force as well as the random force, will obviously modify the random traces, including their scale invariance, isotropicity, etc. We call them the scale-invariant non-Loewnerian forces.  An important class of random forces (here $\xi_t$'s) are those for which their variance behaves like a power-law with time, {\it i.e.}, $\left\langle \xi(t)^2\right\rangle = ct^a$ where $a$ is a positive parameter in the interval $(0,2)$ (one retrieves the Loewner force by setting $a=1$ and $c=\kappa$). An example is presented in Ref.~\cite{credidio2016stochastic} where, after applying the chordal SLE to random traces of anisotropic systems, the driving function was shown to be a 1D time series with power-law variance, the Hurst exponent of which depends on the strength of anisotropy. An other example can be found in Refs.~\cite{oikonomou2008global,nezhadhaghighi2011first}, where the driving function of the ordinary Loewner differential equation contains L\'evy flights in addition to the 1DBM.\\
An important class of 1D correlated time series with power-law variance is the fractional Brownian motion (FBM), for which $a=2H$, where $H$ is the Hurst exponent. For this random force, the Markov property and the stationarity are violated~\cite{reed1995spectral,huy2003remark}. Note that when we have domain Markov property, stationarity and continuity for a time series, then it should be proportional to a one dimensional Brownian motion. FBM is a generalization of the Brownian motion defined by 
\begin{equation}
B_H(t) = \frac{1}{{\Gamma (H + \frac{1}{2})}}\int\limits_0^t {{{(t - s)}^{H - \frac{1}{2}}}} dB(s), 
\label{Eq:FBM} 
\end{equation} 
where $\Gamma$ is the gamma function, $\text{d}B(s)\equiv B(s+\text{d}s)-B(s)$ is the increment of the 1DBM, and the Hurst exponent $H$ is a real number in $(0,1)$. It has the following covariance
\begin{equation} 
\left\langle { {{B_H}(t){B_H}(s)} } \right\rangle = \frac{1}{2}\left( {{{\left| t \right|}^{2H}} + {{\left| s \right|}^{2H}}-{{\left| {t - s} \right|}^{2H}}} \right),
\end{equation}
and also $\left\langle {{B_H}(t)} \right\rangle=0$, where $\left\langle { } \right\rangle$ is the expectation value. A 1DBM is retrieved by setting $H=\frac{1}{2}$, whereas for $H>\frac{1}{2}$ ($H<\frac{1}{2} $) the increments are positively (negatively) correlated, also called the superdiffusive (subdiffusive) regime. The corresponding increments , $ X(t) = {B_H}(t + 1) - {B_H}(t) $, are known to constitute the so-called fractional Gaussian noise. \\

Here we model the random force by a FBM, {\it i.e.}, $\xi=f_R(t)=\xi^{\text{FBM}}\equiv \kappa^H B_H(t)$, so that \[\xi^{\text{FBM}}(\lambda^{1/H}t)\stackrel{d}{=} \lambda\xi^{\text{FBM}}(t),\] and $\left\langle \xi_t^{\text{FBM}}\right\rangle=0$ (note that in the Loewnerian case we have $f_R(t)= \kappa^{\frac{1}{2}} B_t$, as expected). We treat $\kappa$ as an  input parameter, which determines the phase of the trace in the $H=\frac{1}{2}$ limit. To be consistent with the above construction, we should set $a=2H$. If we define $T\equiv t^a$, then $T\rightarrow \lambda^2T$ implies $t\rightarrow \lambda^{2/a}t$, so that $\left\langle {\xi_t^{\text{FBM}}}^2\right\rangle\rightarrow \lambda^2 \left\langle {\xi_t^{\text{FBM}}}^2\right\rangle$.\\

\begin{figure*}
	\includegraphics[scale=0.27]{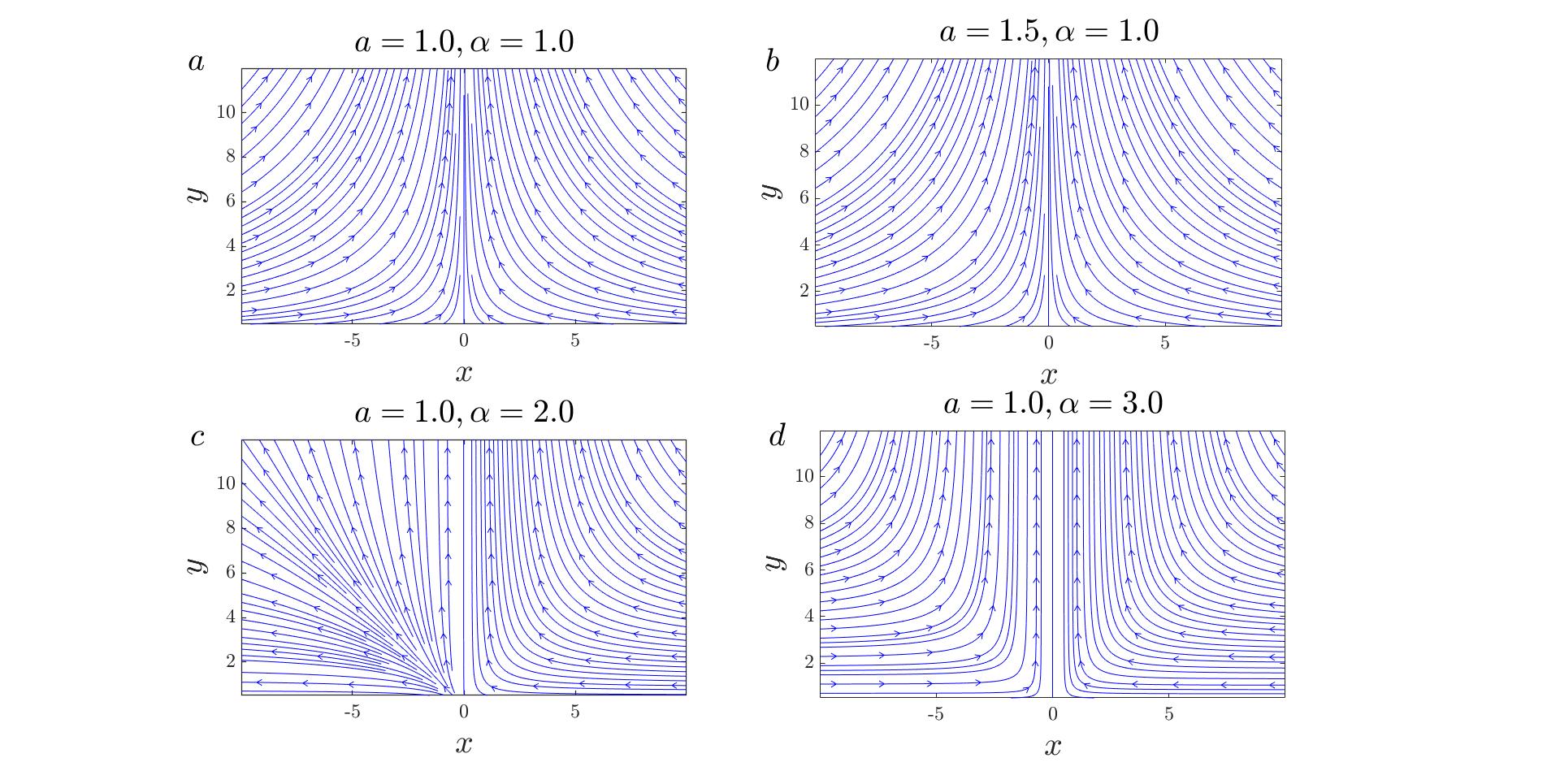}
	\caption{The streamlines of the deterministic forces for different values of $a=2H$ and $\alpha$ (see Eq.~(\ref{Eq:generalizedSLE0})).}
	\label{fig:deterministicForce}
\end{figure*}
Let us now turn to the aim of this paper, {\it i.e.}, to make the exploration process self-similar. Once the FBM random force is incorporated into the evolution Eq.~(\ref{Eq:SLElangavin}), one can easily check that it is impossible to have scale invariance for $a\ne 1$, unless the deterministic force is modified as well. One needs to find the possible forms of the deterministic forces, which also modify the spatial properties of the resulting traces. The next task will then be to obtain the behavior of the traces, including their presumable fractal properties, which will be the focus of the second part of the paper.\\

To make the equation scale invariant, let us multiply the forces by $V_i(r)$, {\it i.e.}, $f_d^i\rightarrow V_i(r)f_d^i$, where $i=x,y$. When we insert this into the evolution equations, and find that
\begin{equation}
\begin{split}
& \frac{\text{d}}{\text{d}t}x_t=-\frac{2x_t}{x_t^2+y_t^2}V_x(x_t,y_t)-\frac{\text{d}}{\text{d}t}\xi_t^{\text{FBM}}\\
& \frac{\text{d}}{\text{d}t}y_t=\frac{2y_t}{x_t^2+y_t^2}V_y(x_t,y_t).
\end{split}
\label{Eq:generalizedSLE0}
\end{equation}
Generally $V_i$ ($i=x,y$) can depend on fixed reference points in space which could break the translational invariance of the system. To preserve translational invariance, these functions should be a power law in terms of $r$, if the only reference point is the origin (which is equivalent to the point $\xi_t$ in ordinary SLE equation). Therefore, we propose the following general form for the deterministic force
\begin{equation}
V_x(x,y)=\frac{x^{\alpha_x-1}}{r^{\beta_x-2}}\ \text{and}\ V_y(x,y)=\frac{y^{\alpha_y-1}}{r^{\beta_y-2}}.
\label{Eq:forces}
\end{equation}
Note that the limit $\alpha_i\rightarrow 1,\beta_i\rightarrow 2$ ($i=x,y$) recovers the Loewner force. Also, $\alpha_x$ should be an integer, so that the function stays analytical in the region $x<0$. We notice that a proportionality constant can be incorporated in Eq.~(\ref{Eq:forces}), that is not of importance, since it can be absorbed by a reparameterization of time, and to be compatible with the limit $a\rightarrow 1$, we set it to unity. The singularity at $r\rightarrow 0$, like the singularity of the Loewner force, is not of central importance. By tuning the parameters for making the process scale invariant, we note that if $\alpha_x\ne \alpha_y$, or $\beta_x\ne\beta_y$, then the system will not be scale invariant because of the dependence on $r=\sqrt{x^2+y^2}$. Therefore we must have $\alpha_x= \alpha_y\equiv\alpha$, and $\beta_x=\beta_y\equiv\beta$, for which $x$ and $y$ have the same scaling behavior. By applying the scale transformation, one can easily show that the relation between the constants is
\begin{equation}
\beta=\alpha-1+\frac{2}{a}.
\label{Eq:alpha-beta-a}
\end{equation}
The trajectories of the tip of the traces corresponding to the forces of Eq.~(\ref{Eq:forces}) for the parameters $a=1$ and $1.5$, and also $\alpha=1,2$ and $3$ are shown in Fig.~\ref{fig:deterministicForce}. These trajectories identify the temporal evolution of trial points in the upper half plane, according to Eq.~(\ref{Eq:generalizedSLE0}) without stochastic force. As is seen in Fig.~\ref{fig:deterministicForce} for $\alpha=1$ (whatever $a$ is) the force field is symmetric ($V_x=V_y$, see Eq.~(\ref{Eq:generalizedSLE0})) and the orientation of the forces is the same as ordinary SLE ($\alpha=1, a=1$). The magnitude of the forces however changes with $a$. For the case $\alpha \ne 1$ ({\it i.e.} $V_x \ne V_y$) both orientations and magnitudes change. For even values of $\alpha$, the structure of the force field changes qualitatively (having the attractor $x=-\infty$ and $y=+\infty$ for $x<0$, and $x=0$ and $y=+\infty$ for $x\geq 0$, in contrast to the ordinary chordal SLE). For odd values of $\alpha$, although the structures are the same as for $\alpha = 1$, the orientation of the stream flows changes. Since the strategy of this study is to minimally change the Loewner evolution equations, we choose the case $\alpha=1$. Note also that the quantity $A$ defined via the differential equation $\text{d}A\equiv \vec{f}_d.\text{d}\vec{l}$ is not a complete differential unless $\alpha=1$, suggesting that there is an extra conservation law in this case, that is not fulfilled for other $\alpha$ values. Therefore, we set $\alpha=1$ throughout the rest of this paper, for which Eq.~(\ref{Eq:SLElangavin}) is modified to
\begin{equation}
\begin{split}
& \frac{\text{d}}{\text{d}t}x_t=-\frac{2x_t}{\left( x_t^2+y_t^2\right)^{\frac{1}{a}}}-\frac{\text{d}}{\text{d}t}\xi_t^{\text{FBM}}\\
& \frac{\text{d}}{\text{d}t}y_t=\frac{2y_t}{\left( x_t^2+y_t^2\right)^{\frac{1}{a}}} .
\end{split}
\label{Eq:generalizedSLE}
\end{equation}
Note also that the deterministic force in the Loewner equation decays with $1/r$ which is long-range. For the subdiffusive FBM ($\beta>2$) this force decays faster than $1/r$, while for the superdiffusive ($\beta<2$) it decays slower.\\

\begin{figure*}
	\includegraphics[scale=0.6]{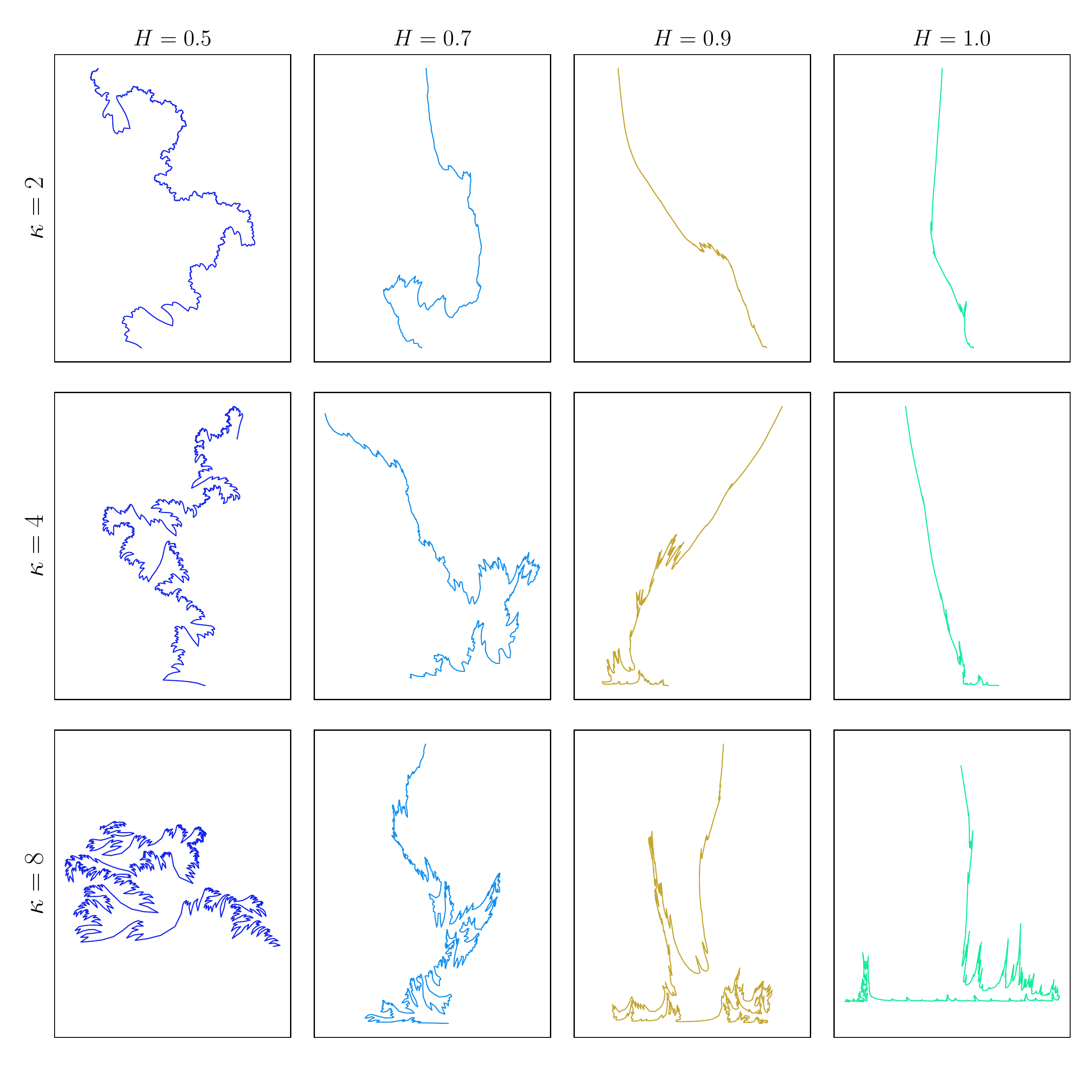}
	\caption{Samples of traces obtained from by Eq.~(\ref{Eq:main}). $H=0.5$ (first column) shows the results for the ordinary SLE traces.}
	\label{fig:boat1}
\end{figure*}
Let us highlight some points about the equation governing $g_t^{\text{NL}}$ (NL stands for non-Loewnerian) for the Eq.~(\ref{Eq:generalizedSLE}) generated by the modified force. We recall that in the ordinary Loewner evolution, the SLE equations are obtained using the fact that $ I_t(w = 0) - \xi_t = x_t + i y_t$. Similarly we define the map $ {I^{\text{NL}}_t}(w = 0) - {\xi_t^{\text{FBM}}} = {x_t} + i{y_t} $, so that, 
\begin{equation}
\frac{\text{d}}{{\text{d}t}}({x_t} + i{y_t}) = \frac{{ - 2({x_t} - i{y_t})}}{{{{({x^2} + {y^2})}^{\frac{1}{a}}}}} - \frac{d}{{dt}}{\xi_t^{\text{FBM}}}, 
\label{Eq:one}
\end{equation}
or equivalently, 
\begin{equation}
\begin{split}
\frac{\text{d}I^{\text{NL}}}{{\text{d}t}} &= \frac{{ - 2}}{{{{({x_t} - i{y_t})}^{\frac{1}{a} - 1}}{{({x_t} + i{y_t})}^{\frac{1}{a}}}}}\\ 
&=\frac{{ - 2}}{{{{({{I^{\text{NL}}}^* } - {\xi_t^{\text{FBM}}})}^{\frac{1}{a} - 1}}{{(I^{\text{NL}} - {\xi_t^{\text{FBM}}})}^{\frac{1}{a}}}}},
\label{Eq:two}
\end{split}
\end{equation}
where the superscript ${}^*$ denotes the complex conjugation operation. Therefore, the equation governing $I^{\text{NL}}_t$ for the modified Loewner evolution is
\begin{equation}
\partial_tI^{\text{NL}}(w) = \left| I^{\text{NL}}(w) - \xi_t^{\text{FBM}} \right|^{2-\frac{2}{a}}\times \frac{{ - 2}}{{I^{\text{NL}}(w) - {\xi_t^{\text{FBM}}}}},
\label{Eq:three}
\end{equation}
where the first term on the right hand side is the additional term with respect to the ordinary SLE equation. Then the direct equation is obtained simply by the transformation $t\rightarrow -t$, so that
\begin{equation}
{\partial_t}{g^{\text{NL}}_t}(z) = \frac{2}{{{{\left| {g^{\text{NL}} - {\xi_t^{\text{FBM}}}} \right|}^{\frac{2}{a} - 2}}(g^{\text{NL}} - {\xi_t^{\text{FBM}}})}}.
\label{Eq:generalSLE0}
\end{equation}
For a piecewise constant driving function, the solution of the above equation (which corresponds to the slit map in the ordinary SLE) is given in the Appendix, see Eq.~(\ref{Eq:starpm}).

\section{The numerical observables}\label{SEC:num}
In this section, we introduce the observables that are investigated here in terms of $\kappa$, which is treated here as an input parameter. By the transformation $T=t^a$, the non-Loewnerian Eqs.~(\ref{Eq:generalizedSLE}) are obtained as, 
\begin{equation}
\begin{array}{l}
\frac{{dx}}{{dT}} =  {T^{\frac{1}{a} - 1}}\frac{{-2a^{-1}x}}{{{{({x^2} + {y^2})}^{\frac{1}{a}}}}} - \frac{d}{{dT}}\xi^{\text{FBM}}(T)\\
\frac{{dy}}{{dT}} = {T^{\frac{1}{a} - 1}}\frac{{2a^{-1}y}}{{{{({x^2} + {y^2})}^{\frac{1}{a}}}}}.
\end{array}
\label{Eq:main}
\end{equation}
\begin{figure*}
    \centerline{\includegraphics[scale=.33]{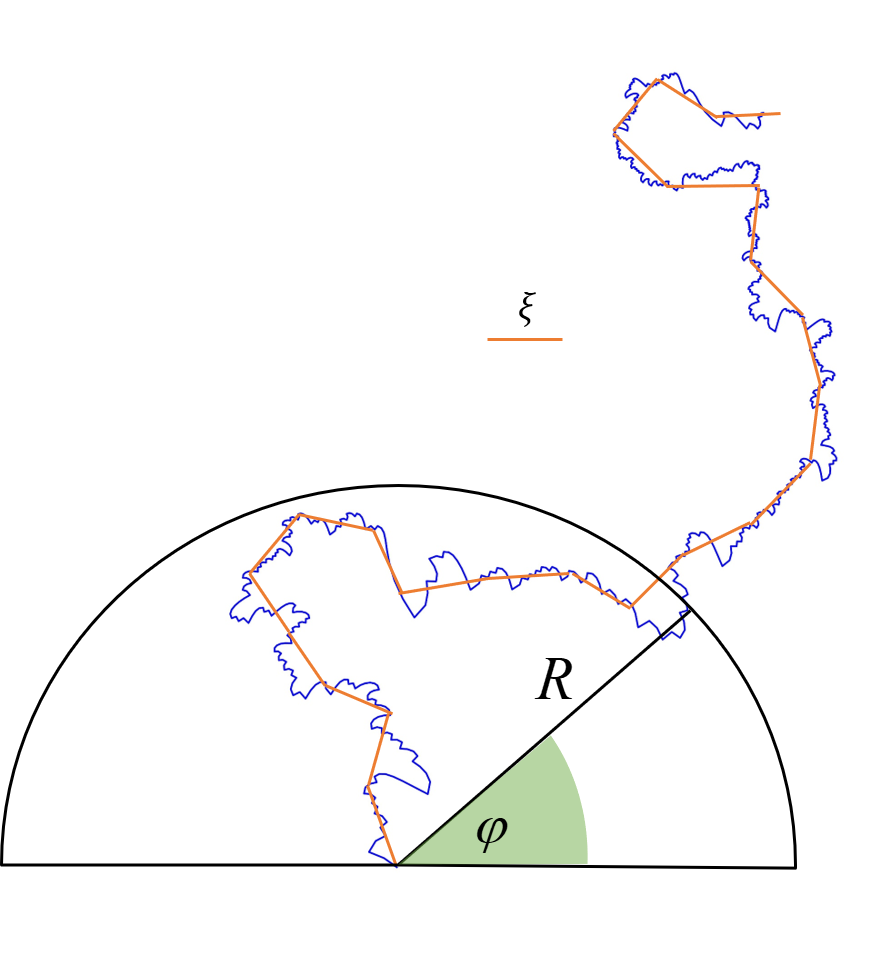}}
	\caption{The schematic procedure of calculating LPP and the fractal dimension using the yardstick method.}
	\label{fig:Scehmatic23}
\end{figure*}
In Fig.~\ref{fig:boat1} we show some examples of traces generated by the above generalized equation for $\kappa=2,4,8$ and $H=0.5, 0.7, 0.9 ,1.0$. We see that, as the Hurst exponent increases, the traces become smoother, and as $\kappa$ increases, the paths become more twisted and rough, leading to higher fractal dimensions. 
To quantify these properties we have applied some standard statistical measures, which are the left passage probability (LPP), the fractal dimension (${D_f}$) and the winding angle statistics. For a curve that goes from the origin to infinity, the LPP corresponds to the probability $p(x,y)$ that the curve passes at the left of the point $(x,y)$. In Fig.~\ref{fig:Scehmatic23} we show the LPP in polar coordinates $(r,\varphi)$. An important test to validate the calculation of the diffusivity parameter is based on Schramm's formula for a chordal $SLE_{k'}$ when applied to systems with conformal invariance, 
\begin{equation}
\begin{split}
p_{\kappa'}(r,\varphi ) = p_{\kappa'}(\varphi )= \frac{1}{2} + &\frac{\Gamma (\frac{4}{\kappa'})}{\sqrt \pi  \Gamma (\frac{8 - \kappa'}{2\kappa' })}\cot (\varphi )\times\\
&{_2}{F_1}(\frac{1}{2},\frac{4}{\kappa' },\frac{3}{2}, - \cot {(\varphi )^2}),
\end{split}
\label{Eq:lpp1}
\end{equation}
where ${}_2F_{1}$ is a hypergeometric function. Also note that $p_{\kappa'}(r,\varphi ) = p_{\kappa'}(\varphi )$ for conformally invariant systems. $\kappa'$ is the test diffusivity parameter and coincides with the input $\kappa$ for the conformally invariant system $H=\frac{1}{2}$. Interestingly, we have observed that, for small enough $H$, this formula remains nearly valid for any $\kappa$ value. Therefore we can define an effective $\kappa_{\text{eff}}$, by minimizing the weighted mean square deviation 
\begin{equation}
K_H(\kappa')=\frac{1}{N}\sum_{i} \sum_{\varphi}[p_H(\varphi,R_i)-p_{\kappa'}(\varphi)]^2,
\label{Eq:l1}
\end{equation}
where $p_H$ is the LPP obtained numerically for curves with Hurst exponent $H$, and the outer sum goes over values of $R_{i+1} =R_i -0.01$ that are computed iteratively for $i=0,1,...,10$  in which $R_0 = \langle R_{\text{max}} \rangle /2$, where  $\langle R_{\text{max}} \rangle$ is defined as the average distance between the beginning and the end of the curves. The second sum goes over $\varphi$ in the interval $0 \leq \varphi \leq \pi$, in steps of $0.01$, and $N$ is the total number of points. The value of $\kappa'$ that minimizes Eq.~(\ref{Eq:l1}) is $\kappa_{\text{eff}}$. \\

Let us define $ R $ as the Euclidean distance between the starting point and the end point of the scale invariant path.  $R$ is related to the length of the curve $l$ via the scaling relation 
\begin{equation}
\sqrt {\left\langle {{R^2}} \right\rangle }  \sim {l^\nu }, 
\label{Eq:nu} 
\end{equation}
where $\left\langle \right\rangle $ denotes the ensemble average, and $\nu$ is related to the fractal dimension via $ \nu  = \frac{1}{{{D_f}}} $. There are also other methods to obtain the fractal dimension, like the yardstick method, as depicted in Fig.~\ref{fig:Scehmatic23}. In this scheme, we use a yardstick of fixed length $ \xi $ to measure the length of the curve which we denote  by $ N(\xi ) $. Following the definition of the capacity dimension, one has $ N(\xi ) \propto {\xi ^{{D_f}}} $ or equivalently, $ {D_f} = \mathop {\lim }\limits_{\xi  \to 0} \frac{{N(\xi )}}{\xi } $. The fractal dimension obtained using this method should be compared with the inverse of $\nu$ to check for consistency.

\section{Simulation results}\label{SEC:results}
\begin{figure*}
	\begin{subfigure}{0.45\textwidth}\includegraphics[width=\textwidth]{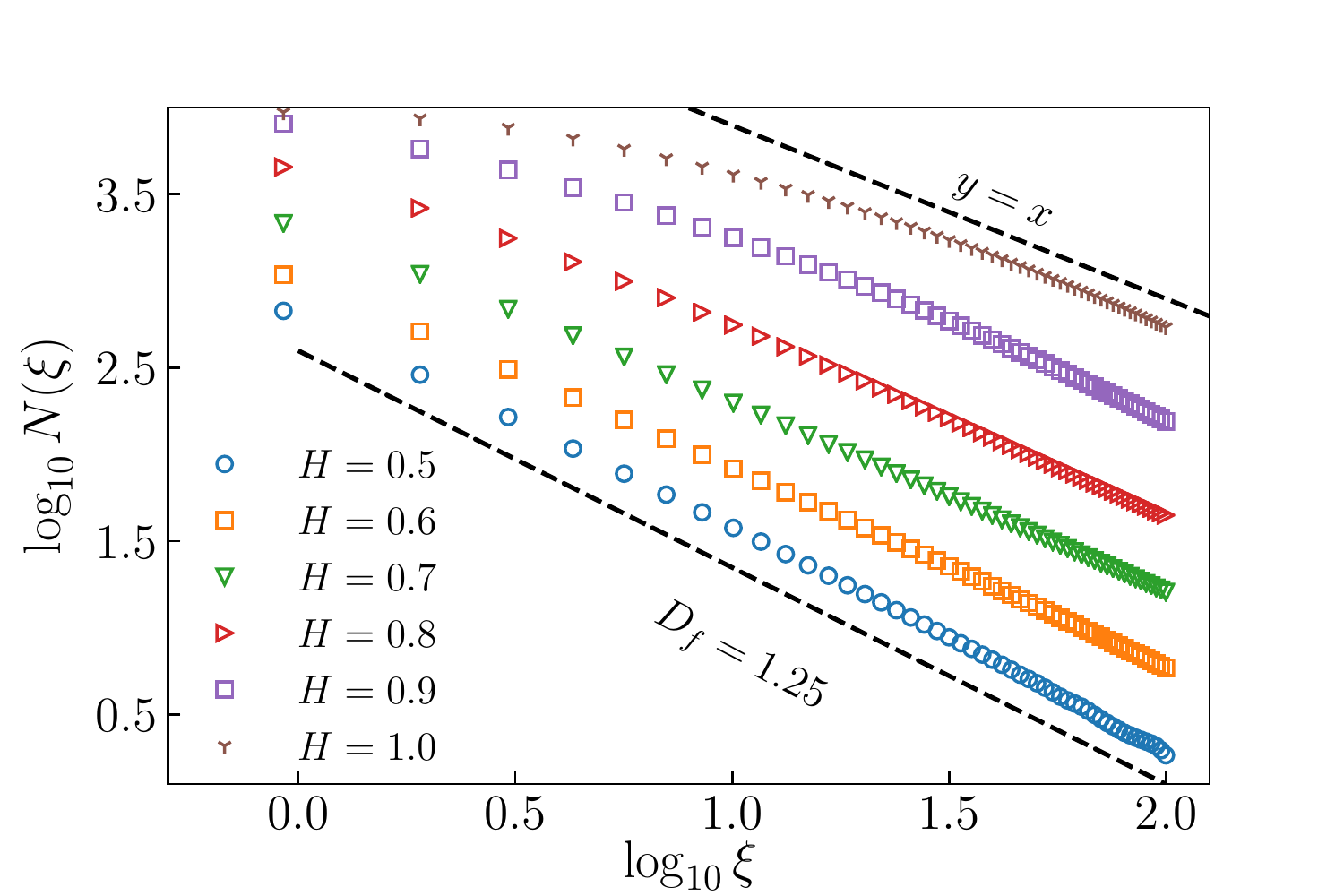}
		\caption{}
		\label{fig:df}
	\end{subfigure}
	\begin{subfigure}{0.45\textwidth}\includegraphics[width=\textwidth]{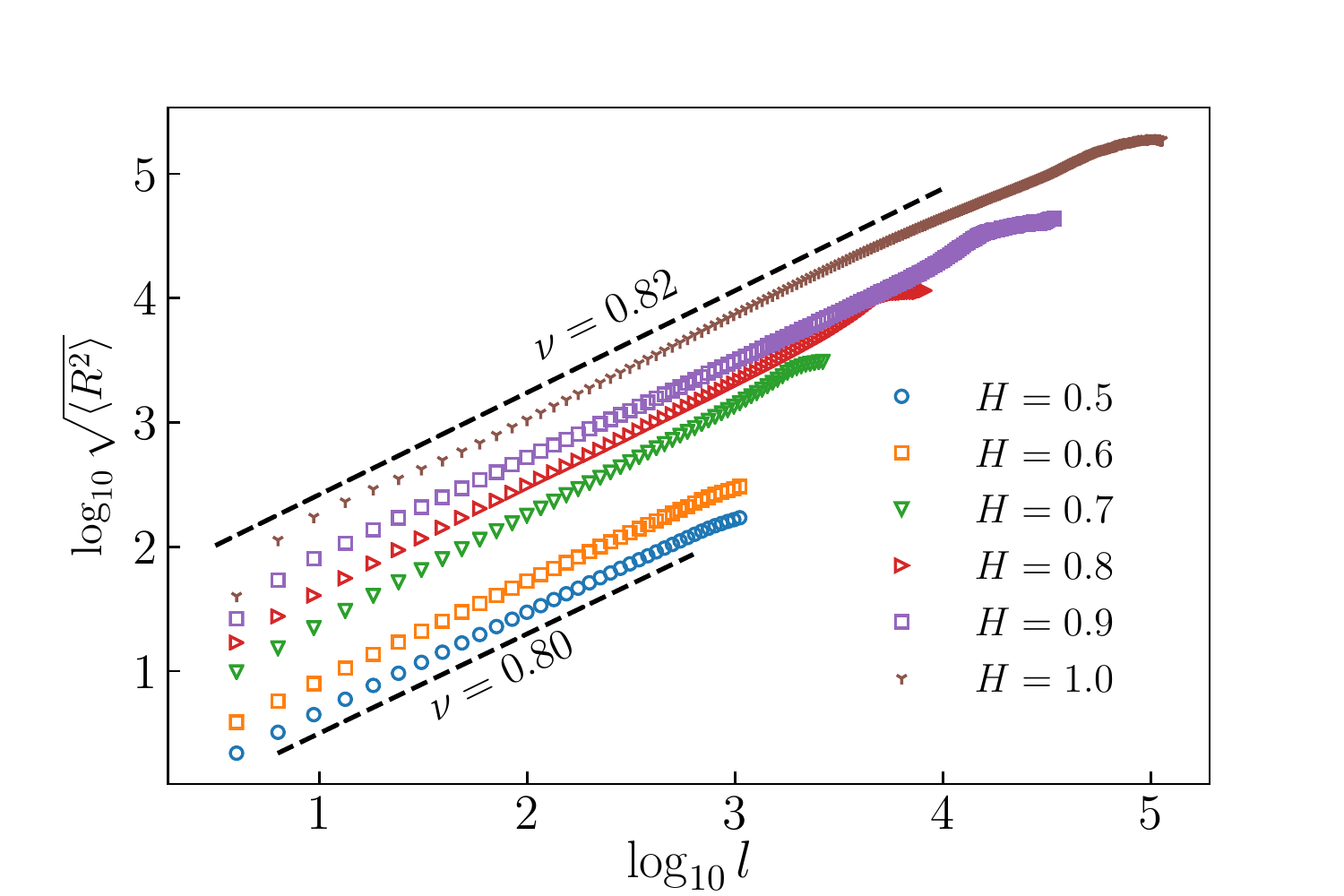}
		\caption{}
		\label{fig:nu}
	\end{subfigure}
  	\caption{(a) Log-log plot of $N$ as a function of $\xi$ for different Hurst exponents $H$ and $\kappa=2$. (b) Log-log plot of $\sqrt {\left\langle {{R^2}} \right\rangle }  $ against $\left\langle l \right\rangle $ for different Hurst exponents for $\kappa=2$.}
	\label{fig:Df_nu}
\end{figure*}

\begin{figure*}
	\begin{subfigure}{0.45\textwidth}\includegraphics[width=\textwidth]{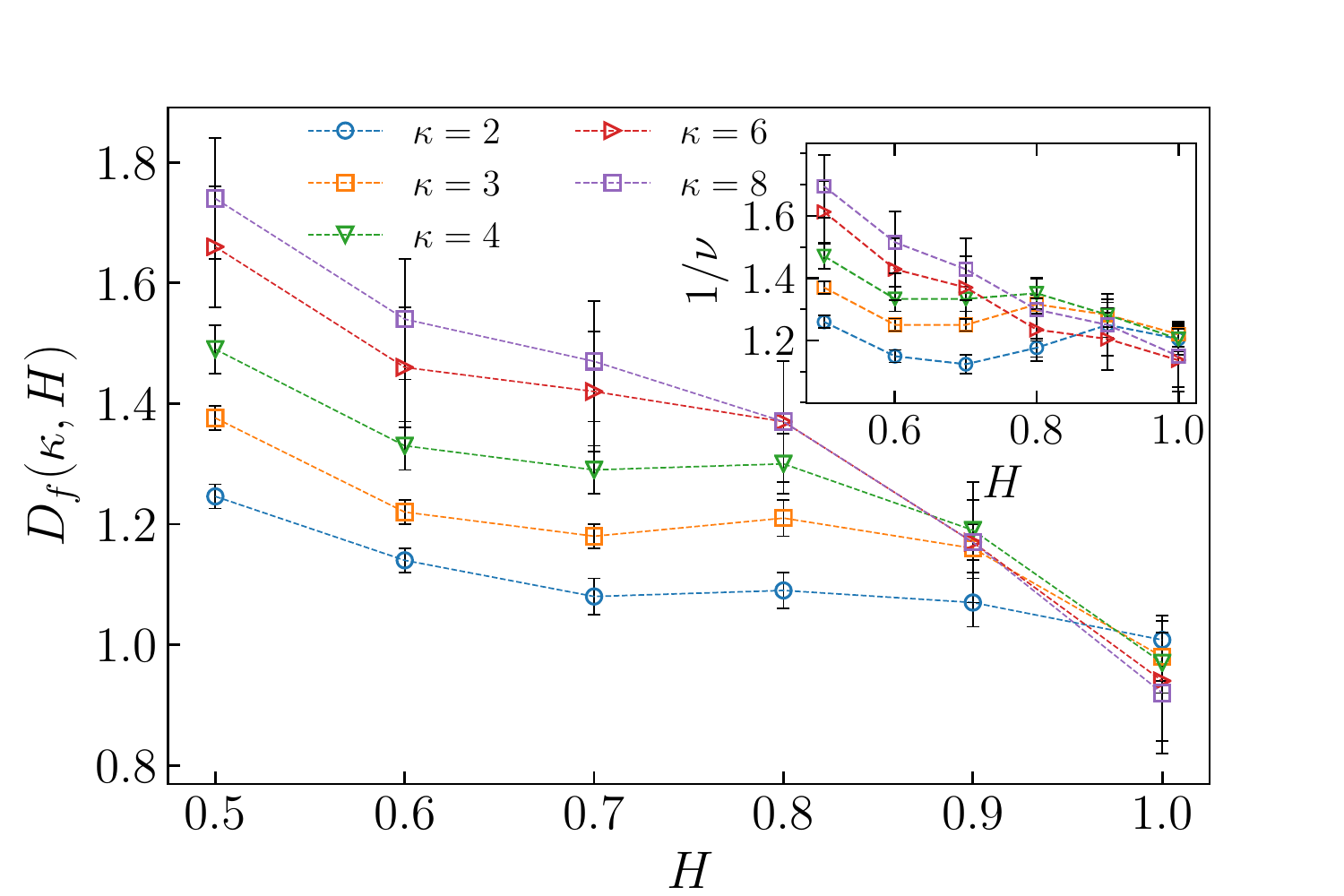}
		\caption{}
		\label{fig:KHdf}
	\end{subfigure}
     \begin{subfigure}{0.45\textwidth}\includegraphics[width=\textwidth]{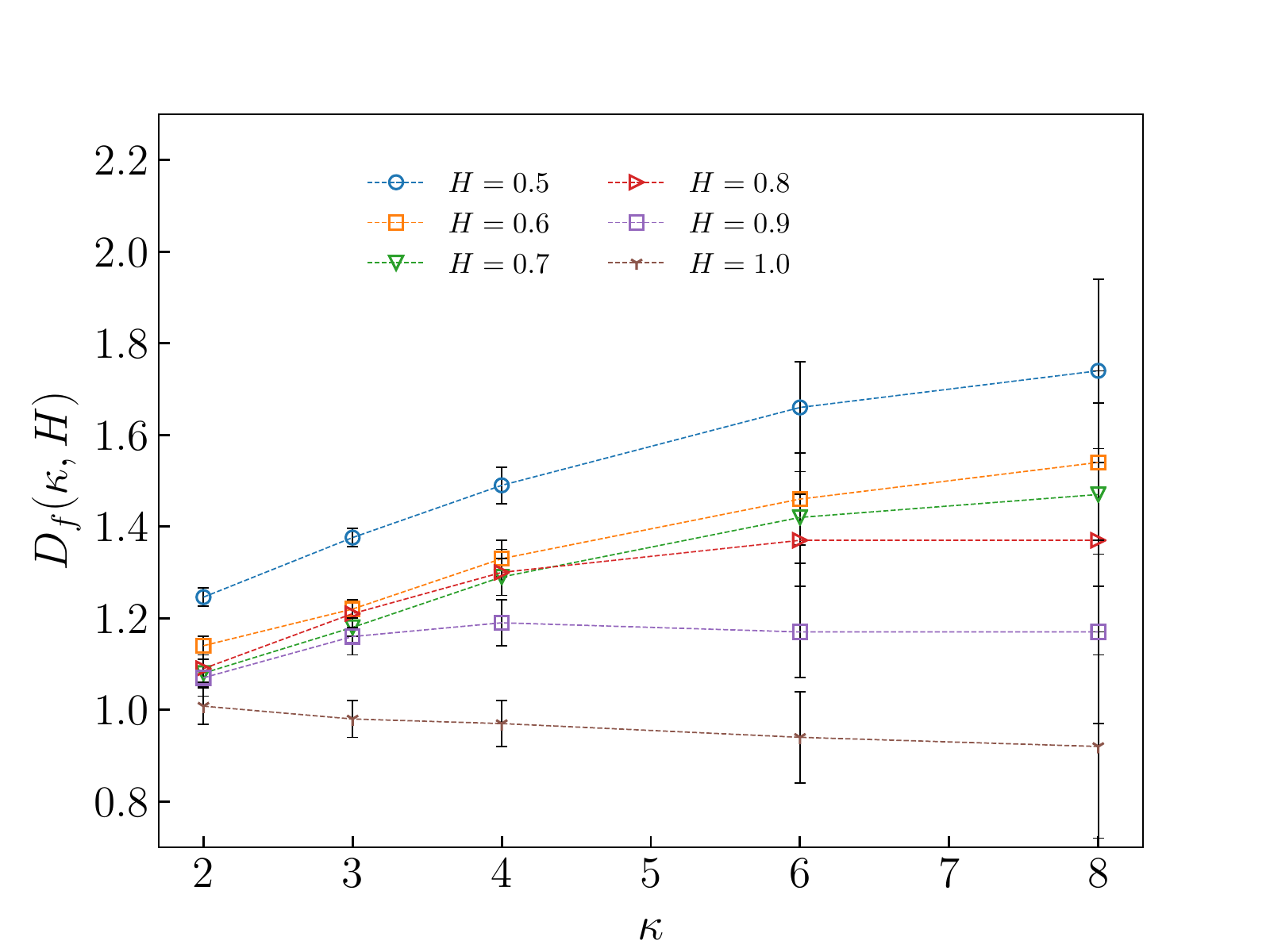}
	\caption{}
	\label{fig:Df_k}
     \end{subfigure}
	\caption{ (a) The fractal dimension ${D_f}(\kappa ,H)$ as a function of $H$ for different diffusivities $\kappa$. Inset: the $\frac{1}{\nu }$ exponent in terms of $H$. (b) The fractal dimension ${D_f}(\kappa ,H)$ as a function of $\kappa$ for various $H$ values.}
	\label{fig:KHnudf}
\end{figure*}

\begin{figure*}
	\includegraphics[scale=0.6]{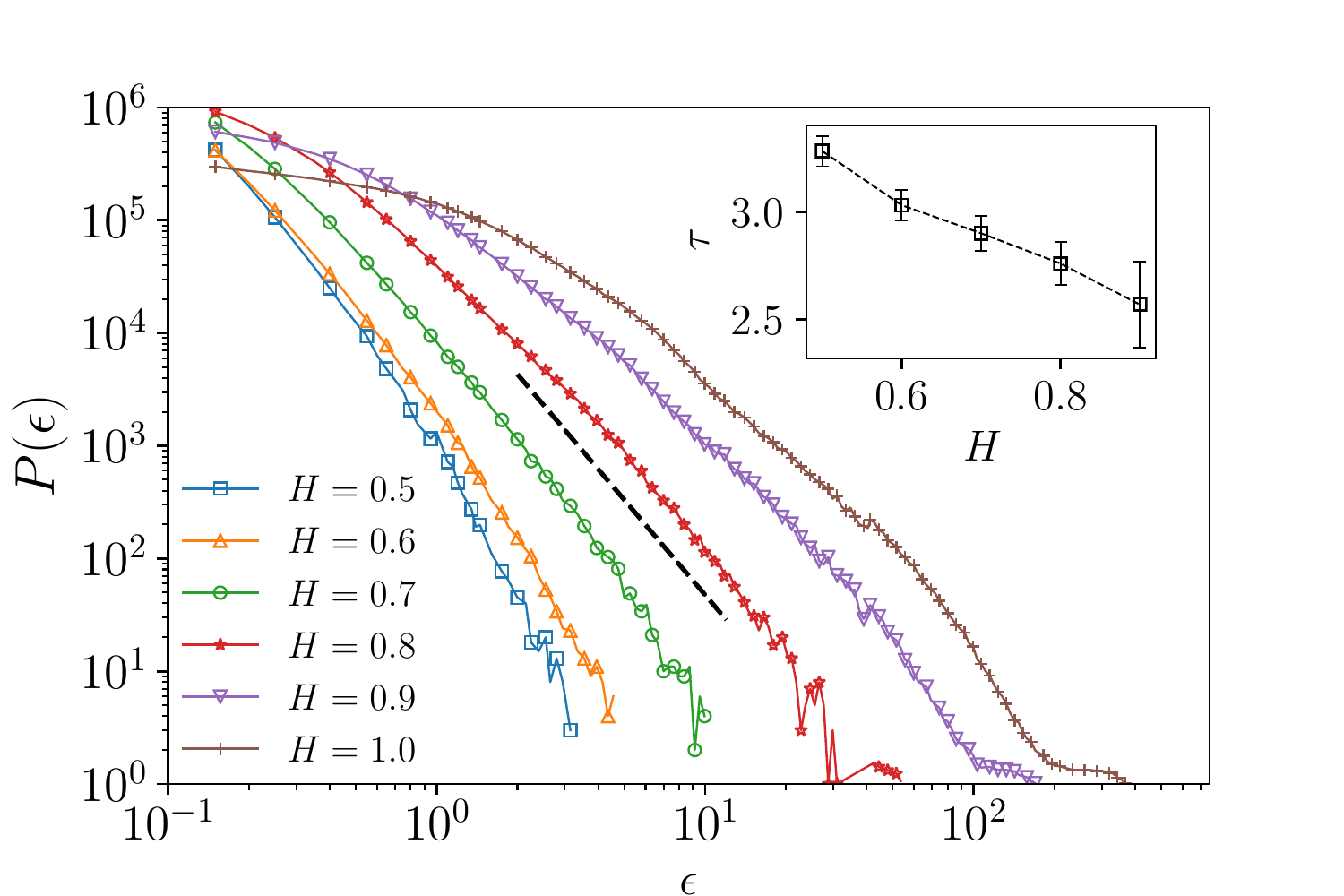}
	\caption{Log-log plot of the distribution function of the distance between two consecutive points ($\epsilon $) in the trajectory for $\kappa=2$ and different Hurst exponents H. Inset: the exponent $\tau$ in terms of $H$.}
	\label{fig:dl}
\end{figure*}

In this section we present the results of the simulations for $H\geq\frac{1}{2}$, setting $t_{\text{max}}=1$ for all samples. We investigate the properties of the model for $\kappa=2, 3, 4, 6, 8$. For the numerical analysis, we produced $4\times10^3$ traces of length $10^4$. To this end, we discretized Eq.~(\ref{Eq:main}) using the finite-difference method for fixed values of $a=2H$ and $\kappa$. To produce the FBM time series, we used the direct method, {\it i.e.}, Eq.~(\ref{Eq:FBM}) for a given $H$. The fractal dimensions obtained with the yardstick method for $\kappa=2$ are presented in Fig.~\ref{fig:df} for $H=0.5, 0.6, 0.7, 0.8, 0.9$ and $1.0$. One should note that $D_{f}=5/4$ for $H=0.5$ ~\cite{najafi2012avalanche}. The analysis for the $\nu$ exponent is presented in Fig.~\ref{fig:nu}. The values for the exponents for $\kappa=2$ can be found in Table~\ref{TAB:exponentsk=2}. We observed power-law behavior for all values of $\kappa$ and $H$, showing that the curves driven by the proposed scale invariant non-Loewnerian forces are self-similar. The results are reported in tables~\ref{TAB:exponentsk=3},~\ref{TAB:exponentsk=4},~\ref{TAB:exponentsk=6}, and~\ref{TAB:exponentsk=8} for $\kappa=3,4,6$ and $8$, respectively. For small Hurst exponents ($H\lesssim 0.7$), the numerical values of $D_{f}$ are in agreement within their error bars  with $\nu_{\kappa}=\frac{1}{D_f^{\kappa}}$. The dependence of  $D_f^{\kappa}$ and $\nu^{-1}_{\kappa}$ on $H$ is presented in Fig.~\ref{fig:KHdf}, from which we see that $D_f$ is a monotonically decreasing function of $H$, as one intuitively expects from the traces presented in Fig.~\ref{fig:boat1}. Note also that the fractal dimension becomes nearly unity as $H\rightarrow 1$ for all $\kappa$ values. The dependence of $D_f^{\kappa}(H)$ on $\kappa$ is shown in Fig.~\ref{fig:Df_k}, displaying a monotonically increasing behavior with average slopes that depend on $H$. As a verification, one can inspect the fractal dimensions for $H=0.5$. The data fits well the relation $D_f=1+\frac{\kappa}{8}$~\cite{cardy2005sle}, which gives $\frac{5}{4}$ for $\kappa=2$, $\frac{11}{8}$ for $\kappa=3$, $\frac{3}{2}$ for $\kappa=4$, $\frac{7}{4}$ for $\kappa=6$ and $2$ (space filling) for $\kappa=8$. The fractal dimensions for $\kappa>4$ show deviations (see Fig.~\ref{fig:Df_k}), which are due to the fact that in this limit the trace touches itself in the thermodynamic limit requiring the generation of a much larger number of numerical samples. In the inset of Fig.~\ref{fig:KHdf} we show the variation of $1/\nu$ which has the same trend as $D_f$, but deviates from it for large values of $H$.\\
One may relate the traces in this limit to L\'evy flights with the scaling relation $\sqrt{\left\langle r^2\right\rangle}\sim l^{1/z} $, where $z$ is called the dynamic exponent ($z=2$ for an ordinary Brownian motion) and is equal to the step index $f$, which is the exponent of the power-law correlated noise in the corresponding Langevin equation~\cite{fogedby1994levy,feller1957introduction}. We suggest that the non-Loewnerian evolution is similar to a L\'evy flight with step index $f=D_f^{\kappa}(H)$. As an example, for $H=0.9$ and $\kappa=4$ we would have $f=1.2\pm 0.1$ for which the upper critical dimension would be $d_c\equiv 2f-2=0.4\pm 0.1$.~\cite{fogedby1994levy} In Fig.~\ref{fig:dl} we plot the distribution function of the distance between two consecutive points ($\epsilon $) for the non-Loewnerian curves for $\kappa=2$. The graphs show power-law behavior for $H=0.6,0.7,0.8$ and $0.9$, while for the uncorrelated case ($H=0.5$, i.e. ordinary SLE) and for the extreme $H=1$ cases where the deviations from the power-law behavior is high. As evidence for the similarity with L\'evy flights, we fit our data with the relation for the distribution function of $\epsilon$, which in two dimensions is~\cite{fogedby1994levy,feller1957introduction}
\begin{equation}
P(\epsilon)d^2\epsilon \propto \epsilon^{-1-f}d\epsilon,
\end{equation}	
or equivalently $P(\epsilon)\propto \epsilon^{-\tau}$, where $\tau=2+f$. The inset of Fig.~\ref{fig:dl} shows the $\tau$ exponent for $\kappa=2$, for which the corresponding $f$ matches the fractal dimension only for small $H$ values. This suggests that, for small enough $H$ values, the system shows similarities with L\'evy flights.\\

\begin{figure*}
	\begin{subfigure}{0.45\textwidth}\includegraphics[width=\textwidth]{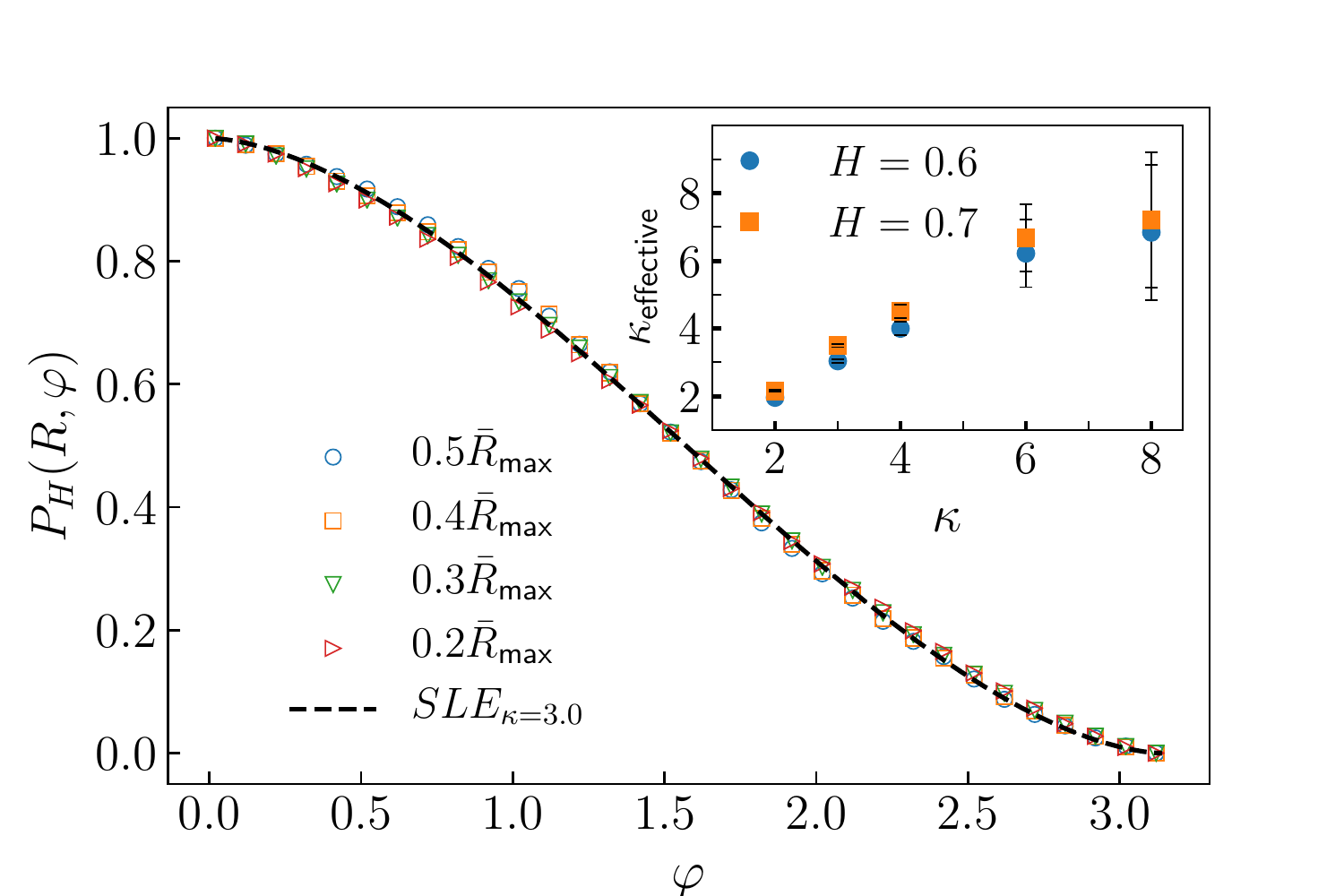}
		\caption{}
		\label{fig:LPP}
	\end{subfigure}
	\begin{subfigure}{0.45\textwidth}\includegraphics[width=\textwidth]{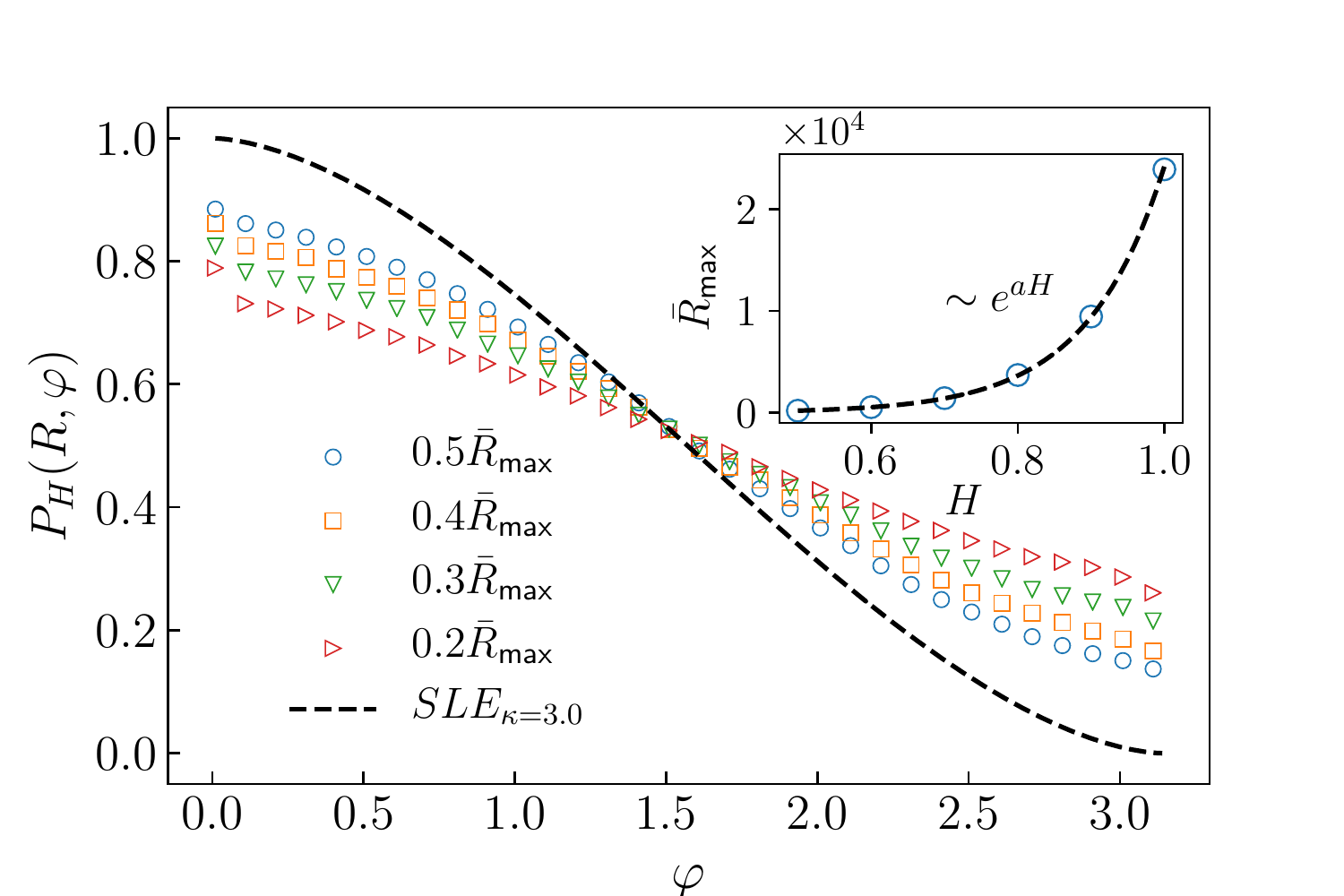}
		\caption{}
		\label{fig:LPPH=1.0}
	\end{subfigure}
\caption{(a) LPP for various radial distances $R$ for $H=0.6$ and $\kappa = 3$. Inset: $\kappa_{\text{eff}}^{\text{LPP}}$ in terms of $\kappa$. (b) LPP for various values of $R$ for $H=1.0$, $\kappa = 3$. Inset: ${\bar R_{\max }}$ in terms of $H$. }
	\label{fig:LPP_kappa}
\end{figure*}

The LPP analysis shows that for small $H$ values (small $\epsilon_H\equiv \frac{H-0.5}{0.5}$) the system is effectively described by the standard SLE with an effective $\kappa$, called $\kappa_{\text{eff}}$, which is obtained from the minimum of $K(\kappa',H)$ in Eq.~(\ref{Eq:l1}). Consider, for example the LPP for $H=0.6$ and $\kappa=3$ which is shown in the main panel of Fig.~\ref{fig:LPP} for various $R<\bar{R}_{\text{max}}$ where $ {{\bar R}_{\max }} = \left\langle {{R_{\max }}} \right\rangle $. We see a nearly perfect matching with the SLE prediction, {\it i.e.}, Eq.~(\ref{Eq:lpp1}) with $\kappa_{\text{eff}}=3.0\pm0.1$. We observe that the LPP fits the SLE prediction of Eq.~(\ref{Eq:lpp1}) quite well for $H<0.8$ with the $\kappa_{\text{eff}}$ that is shown in the inset. As also shown in the inset, the error bars increase with $\kappa$, indicating that the matching becomes worse. For larger Hurst exponents, Eq.~(\ref{Eq:lpp1}) does not work, as depicted in Fig.~\ref{fig:LPPH=1.0} for $H=1$ and $\kappa=3$. We see that in this case $P_H(R,\varphi)$ depends on $R$ showing that it does not obey SLE properties. We report $\kappa_{\text{eff}}$ and the related exponents in Tables~\ref{TAB:exponentsk=2},~\ref{TAB:exponentsk=3},~\ref{TAB:exponentsk=4},~\ref{TAB:exponentsk=6}, and~\ref{TAB:exponentsk=8}. \\

\begin{table*}
	\begin{tabular}{|c | c| c| c|c|c|}
	\hline $H $  & $\nu$  & $D_f$  & $\kappa_{\text{eff}}^{\text{LPP}}$  &$D_f^{\nu}$     &$D_f(\kappa_{\text{eff}}^{\text{LPP}})$  \\
	\hline $0.5$ & $0.795\pm 0.02$ & $1.246\pm 0.02$ & $-$ & $1.26 \pm 0.03$ & $-$  \\
	\hline $0.6$ & $0.868\pm 0.02$ & $1.14\pm 0.02$ & $1.96 \pm 0.03$ & $1.15 \pm 0.03$ & $1.25 \pm 0.004$ \\
	\hline $0.7$ & $0.893\pm 0.03$ & $1.08\pm0.03$ & $2.16 \pm 0.03$ & $1.12\pm 0.04$ & $1.27 \pm 0.004$ \\
	\hline $0.8$ & $0.84\pm 0.03$ & $1.09 \pm0.03$ & $-$ & $1.19\pm 0.04 $ & $-$ \\
	\hline $0.9$ & $0.80\pm 0.03$ & $1.07 \pm 0.04$ & $-$ & $1.25 \pm 0.05$ &$-$ \\
	\hline $1.0$ & $0.82\pm 0.03$ & $1.01 \pm 0.04$ & $-$ & $1.22 \pm 0.04$ &$-$ \\
	\hline
\end{tabular}
	\caption{The exponents $\nu$, $D_f$, $\kappa_{\text{eff}}^{LPP}$  for various Hurst exponents for $\kappa=2$. In the last two columns we show the functions $D_f^{\nu}= \frac{1}{\nu}$ and $D_f(\kappa) = 1+ \frac{\kappa}{8}$ . }
	\label{TAB:exponentsk=2}
\end{table*}
\begin{table*}
	\begin{tabular}{|c | c| c| c|c|c|}
	\hline $H $  & $\nu$  & $D_f$  & $\kappa_{\text{eff}}^{\text{LPP}}$  &$D_f^{\nu}$     &$D_f(\kappa_{\text{eff}}^{\text{LPP}})$  \\
	\hline $0.5$ & $0.73\pm 0.02$ & $1.376\pm 0.02$ & $-$ & $1.37 \pm 0.04$ & $-$  \\
	\hline $0.6$ & $0.8\pm 0.02$ & $1.22\pm 0.02$ & $3.04 \pm 0.05$ & $1.25 \pm 0.04$ & $1.38 \pm 0.006$ \\
	\hline $0.7$ & $0.8\pm 0.02$ & $1.18\pm0.03$ & $3.5 \pm 0.05$ & $1.25\pm 0.04$ & $1.44 \pm 0.006$ \\
	\hline $0.8$ & $0.76\pm 0.02$ & $1.21 \pm0.03$ & $-$ & $1.32\pm 0.03 $ & $-$ \\
	\hline $0.9$ & $0.78\pm 0.03$ & $1.16 \pm 0.04$ & $-$ & $1.28\pm 0.05$ &$-$ \\
	\hline $1.0$ & $0.82\pm 0.03$ & $1.0 \pm 0.05$ & $-$ & $1.22 \pm 0.04$ &$-$ \\
	\hline
\end{tabular}
	\caption{The exponents $\nu$, $D_f$, $\kappa_{\text{eff}}^{LPP}$  for various Hurst exponents for $\kappa=3$. In the last two columns we show the functions $D_f^{\nu}= \frac{1}{\nu}$ and $D_f(\kappa) = 1+ \frac{\kappa}{8}$ . }
	\label{TAB:exponentsk=3}
\end{table*}
\begin{table*}
	\begin{tabular}{|c | c| c| c|c|c|}
	\hline $H $  & $\nu$  & $D_f$  & $\kappa_{\text{eff}}^{\text{LPP}}$  &$D_f^{\nu}$     &$D_f(\kappa_{\text{eff}}^{\text{LPP}})$  \\
	\hline $0.5$ & $0.68\pm 0.03$ & $1.49\pm 0.02$ & $-$ & $1.47 \pm 0.07$ & $-$  \\
	\hline $0.6$ & $0.75\pm 0.03$ & $1.33\pm 0.02$ & $4.0 \pm 0.2$ & $1.33 \pm 0.05$ & $1.5 \pm 0.13$ \\
	\hline $0.7$ & $0.75\pm 0.03$ & $1.29\pm0.03$ & $4.5 \pm 0.2$ & $1.33\pm 0.05$ & $1.56 \pm 0.13$ \\
	\hline $0.8$ & $0.74\pm 0.03$ & $1.3 \pm0.03$ & $-$ & $1.35\pm 0.05 $ & $-$ \\
	\hline $0.9$ & $0.78\pm 0.04$ & $1.19 \pm 0.04$ & $-$ & $1.28 \pm 0.07$ &$-$ \\
	\hline $1.0$ & $0.83\pm 0.04$ & $1.0 \pm 0.1$ & $-$ & $1.20 \pm 0.06$ &$-$ \\
	\hline
\end{tabular}
\caption{The exponents $\nu$, $D_f$, $\kappa_{\text{eff}}^{LPP}$  for various Hurst exponents for $\kappa=4$. In the last two columns we show the functions $D_f^{\nu}= \frac{1}{\nu}$ and $D_f(\kappa) = 1+ \frac{\kappa}{8}$. }
	\label{TAB:exponentsk=4}
\end{table*}
\begin{table*}
	\begin{tabular}{|c | c| c| c|c|c|}
	\hline $H $  & $\nu$  & $D_f$  & $\kappa_{\text{eff}}^{\text{LPP}}$  &$D_f^{\nu}$     &$D_f(\kappa_{\text{eff}}^{\text{LPP}})$  \\
	\hline $0.5$ & $0.64\pm 0.04$ & $1.61\pm 0.1$ & $-$ & $1.56 \pm 0.09$ & $-$  \\
	\hline $0.6$ & $0.7\pm 0.04$ & $1.45\pm 0.1$ & $6.22 \pm 1$ & $1.43 \pm 0.08$ & $1.78 \pm 0.13$ \\
	\hline $0.7$ & $0.73\pm 0.04$ & $1.4\pm0.1$ & $6.68 \pm 1$ & $1.37\pm 0.07$ & $1.86 \pm 0.13$ \\
	\hline $0.8$ & $0.81\pm 0.06$ & $1.35 \pm0.1$ & $-$ & $1.23\pm 0.09 $ & $-$ \\
	\hline $0.9$ & $0.83\pm 0.06$ & $1.20 \pm 0.1$ & $-$ & $1.2 \pm 0.09$ &$-$ \\
	\hline $1.0$ & $0.88\pm 0.06$ & $1.0 \pm 0.1$ & $-$ & $1.14 \pm 0.08$ &$-$ \\
	\hline
\end{tabular}
\caption{The exponents $\nu$, $D_f$, $\kappa_{\text{eff}}^{LPP}$  for various Hurst exponents for $\kappa=6$. In the last two columns we show the functions $D_f^{\nu}= \frac{1}{\nu}$ and $D_f(\kappa) = 1+ \frac{\kappa}{8}$. }
	\label{TAB:exponentsk=6}
\end{table*}
\begin{table*}
	\begin{tabular}{|c | c| c| c|c|c|}
	\hline $H $  & $\nu$  & $D_f$  & $\kappa_{\text{eff}}^{\text{LPP}}$  &$D_f^{\nu}$     &$D_f(\kappa_{\text{eff}}^{\text{LPP}})$  \\
	\hline $0.5$ & $0.59\pm 0.05$ & $1.70\pm 0.1$ & $-$ & $1.7 \pm 0.14$ & $-$  \\
	\hline $0.6$ & $0.66\pm 0.05$ & $1.53\pm 0.1$ & $6.84 \pm 2$ & $1.52 \pm 0.11$ & $1.86 \pm 0.25$ \\
	\hline $0.7$ & $0.71\pm 0.05$ & $1.47\pm0.1$ & $7.2 \pm 2$ & $1.41\pm 0.1$ & $1.9 \pm 0.25$ \\
	\hline $0.8$ & $0.77\pm 0.08$ & $1.37 \pm0.1$ & $-$ & $1.3\pm 0.13 $ & $-$ \\
	\hline $0.9$ & $0.8\pm 0.08$ & $1.17 \pm 0.1$ & $-$ & $1.11 \pm 0.12$ &$-$ \\
	\hline $1.0$ & $0.87\pm 0.08$ & $1.0 \pm 0.1$ & $-$ & $1.15 \pm 0.1$ &$-$ \\
	\hline
\end{tabular}
	\caption{The exponents $\nu$, $D_f$, $\kappa_{\text{eff}}^{LPP}$  for various Hurst exponents for $\kappa=8$. In the last two columns we show the functions $D_f^{\nu}= \frac{1}{\nu}$ and $D_f(\kappa) = 1+ \frac{\kappa}{8}$. }
	\label{TAB:exponentsk=8}
\end{table*}
\section{Conclusions}\label{SEC:conclusion}

This paper was devoted to a generalization of the Loewner forces in the Loewner evolution process. The Loewner force has two components: a deterministic and a random force. We argued that, for the processes in which the random force follows a power-law with time, one can retrieve scale invariance (self-similar traces) by a modification of the deterministic force. We considered a minimal change to define a modified power-law force. We modeled the random force by a fractional Brownian motion (FBM) characterized by the Hurst exponent $H$, which controls the correlations in the model, and becomes Loewnerian when $H=0.5$. \\

The model was then investigated by simulating the random traces that are driven by non-Loewnerian forces. We numerically showed that the traces are self-similar by analyzing the fractal dimensions with two different methods: scaling of the end-to-end distance, and the yardstick method. The fractal dimensions where shown to decrease monotonically with $H$ reaching $D_f=1$ for $H=1$ for all  $\kappa$ values.Finally, the analysis of the left passage probability (LPP) revealed that for small $H$ values (small $\epsilon_H$, especially for $H=0.6$ and $H=0.7$ in this paper) the prediction of SLE, (Eq.~(\ref{Eq:lpp1})), is still applicable. For these values of $H$ we found a new set of effective coefficients $\kappa_{\text{eff}}$.
\appendix
\section{Discretization of the process; piecewise constant driving function}\label{SEC:varying}
\label{app1}

In this Appendix we aim to find the solution of Eq.~(\ref{Eq:generalSLE0}) for a piecewise constant $\xi^{\text{FBM}}$, {\it i.e.}, considering the driving function to be $\xi_i=$constant in an interval $[t_i,t_{i+1}]$. It is more convenient to work with $\eta(T)\equiv\xi^{\text{FBM}}(t)$, and $ \eta(\lambda^2T)\stackrel{d}{=} \lambda\eta(T) $, for which Eq.~(\ref{Eq:generalSLE0}) becomes

\begin{equation}
{\partial _T}{G_T}(z) = \frac{{2{a^{ - 1}}{T^{\frac{1}{a} - 1}}}}{{{{\left| {G_T(z) - \eta (T)} \right|}^{\frac{2}{a} - 2}}(G_T(z) - \eta (T))}},
\label{Eq:generalSLE}
\end{equation}
where we have defined $G_T(z)\equiv g_t(z)$ with the property $G_{\lambda^2}(\lambda z)=\lambda G_T(z)$. Now we suppose that $\eta$ is (piecewise) constant $\eta$ in the time interval $[T_i,T_{i+1}]$. For the ordinary SLE, the solution is $g_{t_{i+1}}(z)=\xi_i+\sqrt{(g_{t_i}(z)-\xi_i)^2+4\delta t_{i}}$. Then by requiring $g_{\delta t_i}(z_1)=\xi_i$ (in which $z_1$ is the first point in the discrete random sequence which should be mapped to $\xi$) we find that $\delta t_i=\frac{1}{4}\left(\text{Im}\left[z_1\right] \right)^2$, and $\xi_i=\text{Re}\left[z_1\right]$, which is called \textbf{slit map}. Now we do the same for the generalized SLE map. For constant $\eta$ we have
\begin{equation}
{\partial _T}{G_T}(z) = \frac{{2{a^{ - 1}}{T^{\frac{1}{a} - 1}}}}{{{{\left| {G(z) - {\eta _i}} \right|}^{\frac{2}{a} - 2}}(G(z) - {\eta _i})}},
\end{equation}
which, after integrating on both sides becomes
\begin{equation}
\int_{G_{T_i}}^{G_{T_{i+1}}} \text{d}G_T {\left| {G(z) - {\eta _i}} \right|^{\frac{2}{a} - 2}}(G(z) - {\eta _i}) = 2a^{- 1}\int_{T_i}^{T_{i+1}} \text{d}T {T^{\frac{1}{a} - 1}}.
\label{Eq:star}
\end{equation}
The right hand side is $2(T_{i + 1}^{\frac{1}{a}} - T_i^{\frac{1}{a}})$, whereas the left hand side of Eq.~(\ref{Eq:star}) has two (real and imaginary ) parts, and setting the imaginary part to zero gives an extra equation to be self-consistent. Assuming that $G=G_1+iG_2$, the left hand side becomes
\begin{equation}
\begin{array}{l}
\int {(d{G_1} + id{G_2})} \left[ \left( G_1 - \eta _i \right)^2 + G_2^2\right]^{\frac{1}{a} - 1}\left({G_1} - \eta _i  + i{G_2}\right).
\end{array}
\end{equation}
The imaginary part of which is
\begin{equation}
\int\left[ \left( G_1 - \eta _i \right)^2 + G_2^2\right]^{\frac{1}{a} - 1}\left[(G_1-\eta_i)\text{d}G_2+G_2\text{d}G_1 \right] =0.
\end{equation}\\
Since this equation should be satisfied for all trajectories, the argument of the integral should vanish, so that
\begin{equation}
\begin{split}
&({G_1} - \eta _i )d{G_2} + {G_2}d{G_1} = 0\\
&\Rightarrow \frac{{d{G_2}}}{{{G_2}}} =  - \frac{{d{G_1}}}{{{G_1} -\eta _i }}\\
&\Rightarrow {G_2} = {G_2}^0\left( {\frac{{{G_1}^0 - \eta _i }}{{{G_1} - \eta _i }}} \right),
\end{split}
\end{equation}\\
where $G_1^0$ and $G_2^0$ are some constants. Now we apply this to the real part of the integral, giving us
\begin{widetext}
\begin{equation}
\begin{split}
&\int \left[ \left( G_1 - \eta _i \right)^2 + G_2^2\right]^{\frac{1}{a} - 1}\left[({G_1} - \eta _i )\text{d}G_1-G_2\text{d}G_2\right] =\int \left[ \left( G_1 - \eta _i \right)^2 + G_2^2\right]^{\frac{1}{a}}\frac{\text{d}G_1}{G_1-\eta_i}\\
&=-\frac{a}{2}\left(\frac{(G_1-\eta_i)^2}{(G_1^i-\eta_i)^2{G_2^i}^2} \right)\left((G_1-\eta_i)^2+\frac{(G_1^i-\eta_i)^2{G_2^i}^2}{(G_1-\eta_i)^2} \right)^{1+\frac{1}{a}}{_2}F_1\left[1,1+\frac{1}{2a},1-\frac{1}{2a},\frac{(G_1-\eta_i)^4}{(G_1^i-\eta_i)^2{G_2^i}^2}\right]. 
\end{split}
\end{equation}
\end{widetext}
Let us define the last line as $F_a\left(G_1,\eta_i\right) $, then one can easily show that
\begin{equation}
\begin{split}
F_a\left(G_1^i,\eta_i\right)=&-\frac{a}{2{G_2^i}^2}\left((G_1^i-\eta_i)^2+{G_2^i}^2\right)^{1+\frac{1}{a}}\times\\
&{_2}F_1\left[1,1+\frac{1}{2a},1-\frac{1}{2a},\frac{(G_1^i-\eta_i)^2}{{G_2^i}^2}\right].  
\end{split}
\end{equation}
Therefore the final result is
\begin{equation}
T_{i + 1}^{\frac{1}{a}}=T_i^{\frac{1}{a}} -\frac{1}{2}\left[F_a\left(G_1^{i+1},\eta_i\right)-F_a\left(G_1^i,\eta_i\right) \right].
\label{Eq:starpm}
\end{equation}

\bibliography{refs}

\end{document}